\documentclass[a4paper,fleqn,usenatbib]{mnras}

\usepackage{txfonts}

\usepackage[T1]{fontenc}
\usepackage{ae,aecompl}

\usepackage{graphicx}	
\usepackage{amssymb}	

\usepackage{CJKutf8}

\newcommand{\iso}[2]{{}^{#2}\mbox{#1}}

\title[Production uncertainties of $p$ nuclides in SNe Ia]
{\LARGE Uncertainties in the production of $p$ nuclides in thermonuclear supernovae determined by Monte Carlo variations}

\author[Nishimura~el~al.]
{
N.~Nishimura \begin{CJK}{UTF8}{ipxm}(西村信哉)\end{CJK}$^{1, 2}$\thanks{e-mail: nobuya.nishimura@yukawa.kyoto-u.ac.jp}\thanks{UK Network for Bridging Disciplines of Galactic Chemical Evolution (BRIDGCE), \url{https://www.bridgce.net}},
  T. Rauscher$^{3, 4}$\footnotemark[2],
  R.~Hirschi$^{2, 5}$\footnotemark[2],
  A.~St.~J.~Murphy$^{6}$\footnotemark[2],\newauthor
  G.~Cescutti$^{4,7}$\footnotemark[2],
  and C. Travaglio$^{8}$\thanks{B2FH Association, c/o Strada Osservatorio 20, I-10023 Torino, Italy}
\\
  $^1$ Yukawa Institute for Theoretical Physics, Kyoto University, Kyoto 606-8502, Japan\\
  $^2$ Astrophysics Group, Faculty of Natural Sciences, Keele University, Keele ST5 5BG, UK\\
  $^3$ Department of Physics, University of Basel, 4056 Basel, Switzerland\\
  $^4$ Centre for Astrophysics Research, University of Hertfordshire, Hatfield AL10 9AB, UK\\
  $^5$ Kavli IPMU (WPI), University of Tokyo, Kashiwa 277-8583, Japan\\
  $^6$ School of Physics and Astronomy, University of Edinburgh, Edinburgh, EH9 3FD, UK\\
  $^7$ INAF, Osservatorio Astronomico di Trieste, I-34131 Trieste, Italy\\
  $^8$ INFN, Sezione di Torino, I-10125 Torino, Italy
}

\date{Accepted XXX. Received YYY; in original form ZZZ}

\pubyear{2017}

\begin{document}
\label{firstpage}
\pagerange{\pageref{firstpage}--\pageref{lastpage}}
\maketitle

\begin{abstract}
Thermonuclear supernovae originating from the explosion of a white dwarf accreting mass from a companion star have been suggested as a site for the production of $p$ nuclides. Such nuclei are produced during the explosion, in layers enriched with seed nuclei coming from prior strong $s$ processing. These seeds are transformed into proton-richer isotopes mainly by photodisintegration reactions. Several thousand trajectories from a 2D explosion model were used in a Monte Carlo approach. Temperature-dependent uncertainties were assigned individually to thousands of rates varied simultaneously in post-processing in an extended nuclear reaction network. The uncertainties in the final nuclear abundances originating from uncertainties in the astrophysical reaction rates were determined. In addition to the 35 classical $p$ nuclides, abundance uncertainties were also determined for the radioactive nuclides $^{92}$Nb, $^{97,98}$Tc, $^{146}$Sm, and for the abundance ratios $Y(\iso{Mo}{92})/Y(\iso{Mo}{94})$, $Y(\iso{Nb}{92})/Y(\iso{Mo}{92})$, $Y(\iso{Tc}{97})/Y(\iso{Ru}{98})$, $Y(\iso{Tc}{98})/Y(\iso{Ru}{98})$, and $Y(\iso{Sm}{146})/Y(\iso{Sm}{144})$, important for Galactic Chemical Evolution studies. Uncertainties found were generally lower than a factor of 2, although most nucleosynthesis flows mainly involve predicted rates with larger uncertainties. The main contribution to the total uncertainties comes from a group of trajectories with high peak density originating from the interior of the exploding white dwarf. The distinction between low-density and high-density trajectories allows more general conclusions to be drawn, also applicable to other simulations of white dwarf explosions.
\end{abstract}

\begin{keywords}
nuclear reactions, nucleosynthesis, abundances -- stars: abundances -- supernovae: general
\end{keywords}



\section{Introduction}
\label{sec:intro}

Thermonuclear supernovae (Type Ia supernovae, SNe Ia) are thought to involve the explosion of a white dwarf (WD) \citep{wheelerrev,nomotopop,branch}. The two most popular evolution mechanisms leading to an explosion are the merging of two WDs \citep[double-degenerate model,][]{ibe84,webbink} or mass accretion from a companion star on to the surface of a WD in a binary system \citep[single-degenerate model,][]{whe73,nom82}. In the single-degenerate case, material from the outer envelope of the companion star is added to the surface of a WD, where it not only increases its mass but also undergoes thermonuclear H- and He-burning. The combined action of mass increase and heating by nuclear burning triggers explosive C/O-burning which completely disrupts the WD. The layer accreted from the companion star before the explosion in the single-degenerate model may become enriched in $s$-process nuclei by additional $s$ processing in He-burning on the surface of the WD during the accretion phase. In the WD explosion, these $s$-process seed nuclei undergo further nuclear reactions, mainly photodisintegrations, and it was suggested that this could lead to the production of proton-rich, stable isotopes, the so-called $p$ nuclides \citep{howmey}. These nuclides are by-passed by the classical neutron capture processes, the $s$ process and the $r$ process \citep{cam,burb}. Although their abundance is low compared to those of other isotopes in the same element, the very existence of $p$ nuclides poses a long-standing puzzle in investigations of nucleosynthesis and Galactic Chemical Evolution \citep[GCE; see, e.g.,][and references therein]{arngor,p-review,pign-rev}.

The original, parametrized calculations for the single-degenerate model, however, found an underproduction of light $p$ nuclides, even when assuming a seed enrichment of factors $10^3-10^4$ in $s$-process nuclei in the accreted layer \citep{howmey,arngor}. Subsequent studies based on the carbon deflagration model of \citet{nom} found similar problems \citep{kusa05}. This moved the focus of studying the origin of $p$ nuclides to their production in the O/Ne shells of exploding massive stars (core-collapse supernovae, ccSNe), where also photodisintegrations act during the passage of the supernova shock \citep{arn,woohow,hoff96}. Also this site, however, cannot produce all $p$ nuclides in solar proportions. Deficiencies were found in the production of the lightest $p$ nuclides and also in the region of nuclear mass numbers $150\leq A \leq165$ \citep{rhhw02}. The interest in thermonuclear supernovae as sources of $p$ nuclides was renewed when \citet{travWD} post-processed tracers of high-resolution 2D models and found that most $p$ nuclides can be co-produced, provided strong enhancements in the assumed $s$-process seeds are present. This was confirmed by \cite{kusa11}, and it was concluded that a high-resolution treatment of the outer zones of the Type Ia supernova is crucial to accurately follow the production of $p$ nuclides. Altough the assumption of the strong seed enhancement originally was ad hoc, recent calculations by Battino~et~al. \citep[in preparation; see also][]{enrich} confirm the possibility of such an enhancement by thermonuclear pulses in the He-burning accretion layer on the surface of a WD.

In \citet{pinccSn}, we derived $p$-nucleus production uncertainties stemming from uncertainties in astrophysical reaction rates in ccSNe in a Monte Carlo post-processing approach. In this work, we apply the same method to study $p$ production in trajectories obtained with tracers in a 2D SNe Ia model. In order to allow cross-comparison, we base our analysis on tracers from the same model that was also used in \citet{travradio} and also include the radioactive nuclides $^{92}$Nb, $^{97,98}$Tc, and $^{146}$Sm in addition to the 35 classical $p$ nuclides. These now extinct radioactivities were present in the early Solar system \citep{p-review,travradio} and detailed knowledge of their production is important in GCE calculations.

The astrophysical model from which the tracers were taken is briefly outlined in Section~\ref{sec:astrotracers}. The method of the Monte Carlo variation of reaction rates is presented in Section \ref{sec:montecarlo}. The results are shown and discussed in Section~\ref{sec:results}, and a summary is given in Section~\ref{sec:summary}.

\section{Methods}
\label{sec:methods}

\subsection{Astrophysical model}
\label{sec:astrotracers}

The trajectories required for post-processing were obtained by recording the temporal evolution of temperature and density of tracer particles followed through the explosion in a 2D simulation of the explosion of a Chandrasekhar-mass WD consisting of C and O \citep[metallicity $Z=0.02$,][]{dominguez}, assuming an admixture of increased $s$ abundances. The simulation starts at the onset of the explosion, which is ignited in multiple sparks as presented by \citet{kasen}. The explosion itself is treated as a delayed detonation, in which the burning front turns from an initial deflagration into a detonation. The calculation here implements the case DDT-a introduced in \citet{travWD}. The numerical treatment using the combustion code \textsc{leafs} \citep{roehi05} has been discussed in detail in \citet{travWD}, and references therein.  A Lagrangian component in the form of 51200 tracer particles of equal mass was overlaid on top of the Eulerian grid of the combustion code. \citet{sei10} argued that this number of tracers provides sufficient resolution for such a 2D case. A subset of the same tracers has been used recently to study the production of $p$ nuclei in SN Ia by \citet{travWD,travradio,trav15}.

\begin{figure}
\includegraphics[width=\columnwidth]{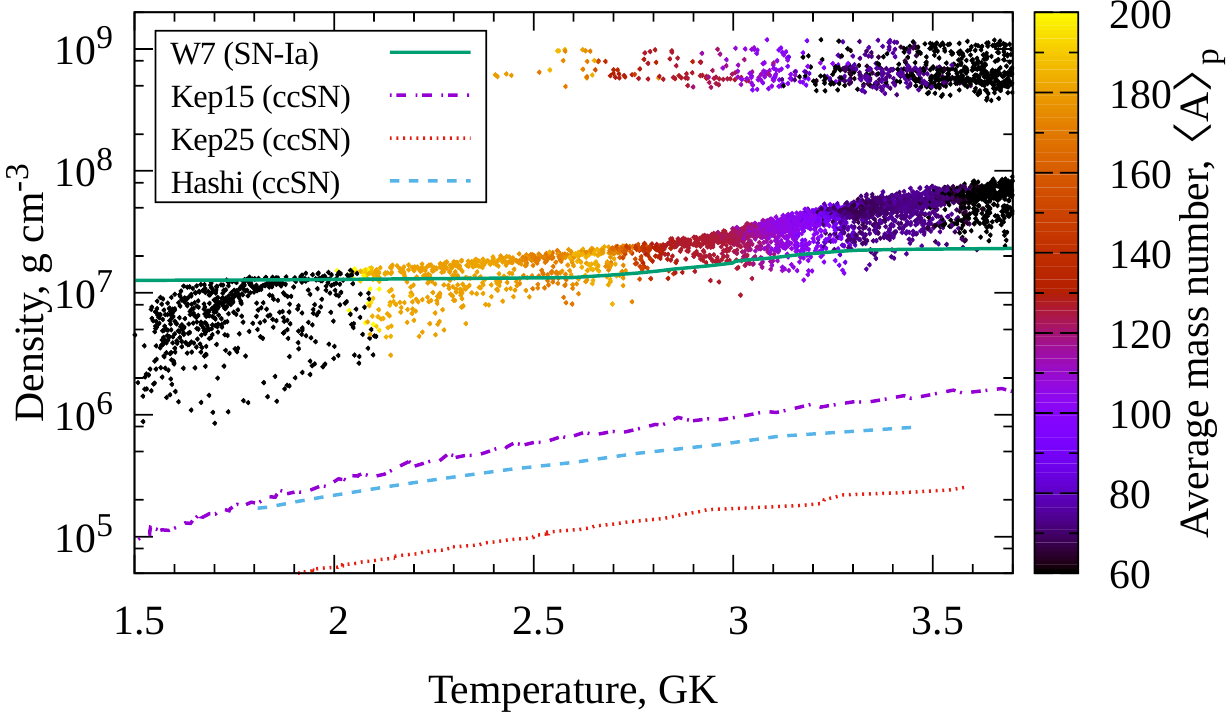}
\caption{Peak density versus peak temperature plotted for each trajectory. The colour of each dot gives the average mass number $\langle A \rangle_\mathrm{p}$ of $p$ nuclei produced in that trajectory. For comparison, density and temperature peak values for four other models are shown: the W7 model for an SN Ia \citep{nom} and the three models of O/Ne layers of exploding massive stars used in \citet{pinccSn}, with one for a 15 $M_\odot$ star (Kep15) and two for 25 $M_\odot$ stars (Kep25, Hashi). \label{fig:2groupsother}}
\end{figure}

Among all tracers, 4624 trajectories were selected to be processed in the MC approach. The remaining tracers experienced peak temperatures which were clearly either too high and thus leading to destruction of all $p$ nuclei or too low and preventing the conversion of the seed distribution into proton-rich nuclei. The selected trajectories show peak temperatures between 1.5 and 3.7 GK. Fig.~\ref{fig:2groupsother} shows the distribution of the peak densities versus peak temperatures for the chosen trajectories. The colour coding provides information on the average mass number of $p$ nuclei $\langle A \rangle _\mathrm{p}$ produced in each trajectory. It is defined as
\begin{equation}
\langle A \rangle_\mathrm{p} = \frac{\sum_{i\in \mathrm{p-nuclei}}{A_i\left(\Delta Y_i/Y_{i,\odot}\right)}}{\sum_{i\in \mathrm{p-nuclei}}{\left(\Delta Y_i/Y_{i,\odot}\right)}} \quad,
\end{equation}
where $Y_i$ is the abundance of a $p$ nuclide, $Y_{i,\odot}$ is the solar abundance of the same nuclide, and $A_i$ is its nuclear mass number. As well known, lighter $p$ nuclei are produced at higher peak temperature whereas heavier $p$ nuclei are produced at lower temperature and only destroyed at high temperature due to their lower binding energy \citep{woohow,p-review}. The figure also shows that the selected trajectories safely encompass the temperature range relevant for $p$ production.

The same initial abundances as in the DDT-a ($Z=0.02$) model of \citet{travWD} were used. This implies an $s$-process distribution peaked at the lighter nuclides with an enhancement between 10 to 1000 compared to solar. We do not expect that the derived uncertainties depend strongly on the pattern of the seed abundances, though, as the main nuclear process is photodisintegration. The photodisintegration path is independent of abundances when it does not have to compete with capture reactions, as long as there are seeds to be photodisintegrated.

\subsection{Monte Carlo variations}
\label{sec:montecarlo}

The selected trajectories were post-processed using the \textsc{pizbuin} code suite, consisting of a fast reaction network and a parallelized Monte Carlo driver. We followed the same procedure as presented in detail in \citet{pinccSn}. Each trajectory was run 10\,000 times in a network calculation, with different rate variation factors, and the combined output was analysed subsequently. This simultaneous variation of rates is superior to a decoupled variation of individual rates as performed in the past because neglecting a combined change in rates may lead to an overemphasis of certain reactions and a misestimate of their impact on the total uncertainty \citep{pinccSn,omeg17}. We define key rates to be those dominating the uncertainty of a given final abundance. By this definition, reducing the uncertainty of a key rate will also considerably decrease the uncertainty in the final abundance. The identification of key rates is achieved by examining the correlation between a change in a rate and the change of an abundance, as found in the stored Monte Carlo data. As before, the Pearson product--moment correlation coefficient \citep{pearson} is used to quantify correlations. Positive values of the Pearson coefficients $r$ indicate a direct correlation between rate change and abundance change, whereas negative values signify an inverse correlation, i.e., the abundance decreases when the rate is increased. The larger the absolute value of the Pearson coefficient, the stronger the correlation. As in \citet{pinccSn,weakNobuya}, a key rate is identified by $|r|\geq 0.65$.

Other correlation definitions have been tested but the Pearson coefficient gave the most reliable results and is simple to handle, especially when calculating a combined, weighted correlation including many trajectories \citep{pinccSn}. Rank correlation methods, although formally assumed to better account for data outliers, are losing too much information in the ranking procedure and are rather unsuited for our purposes when used with reaction rate data \citep{kendall55,mathguru}. Moreover, data outliers to which the Pearson coefficient would be vulnerable do not appear in an analytic variation of reaction rates, anyway.

Each astrophysical reaction rate from Fe to Bi was varied within its own uncertainty range. Forward and reverse rates received the same variation factor as they are connected by detailed balance. The assigned uncertainty range is temperature dependent and constructed from a combination of the measured uncertainty (if available) for target nuclei in their ground states and a theory uncertainty for predicted rates on nuclei in thermally excited states. Theory uncertainties were different depending on the reaction type and can be asymmetric. See \citet{pinccSn,omeg17} for further details.

The employed reaction network contained 1342 nuclides, including nuclides around stability and towards the proton-rich side of the nuclear chart. The standard rate set and the assigned uncertainties were the same as previously used in \cite{pinccSn} and \cite{weakNobuya}: Rates for neutron-, proton-, and $\alpha$-induced reactions were a combination of theoretical values by \citet{adndt00}, supplemented by experimental rates taken from \cite{kadonis} and \cite{cyburt}; decays and electron captures were taken from a REACLIB file compiled by \citet{freiburghaus} and supplemented by rates from \cite{taka} and \cite{gor99} as provided by \cite{NetGen05} and \cite{NetGen13}.

Although the required computational time is independent of how many rates are varied, it depends on the time to complete one network run, the number of MC iterations, and the number of trajectories to be processed. For this present study, the reaction network had to be run at least 40 million times (not counting failed runs). This necessitated the use of HPC facilities, such as the UK DiRAC facility (further facilities used are given in the Acknowledgments).

\section{Results and Discussion}
\label{sec:results}

\subsection{Total uncertainties and key rates}
\label{sec:totalresults}

\begin{figure}
	\includegraphics[width=\columnwidth]{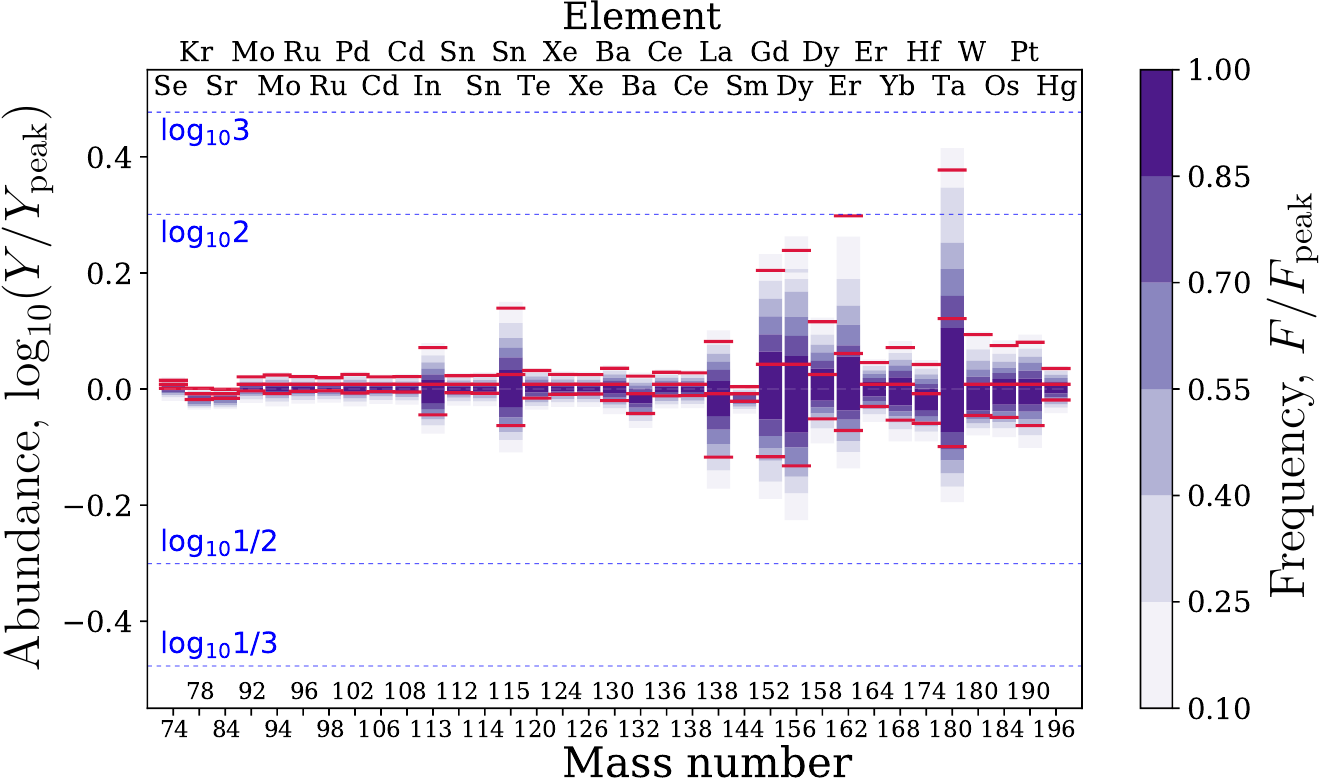}
	\caption{Total production uncertainties of $p$ nuclei due to rate uncertainties in the MC post-processing of the SN Ia model DDT-a. The colour shade gives the relative probabilistic frequency and the horizontal red lines are cumulative frequencies of $5\%$, $50\%$, and $95\%$, enclosing a $90\%$ interval for each nuclide. Uncertainty factors of 2 and 3 are marked by dotted lines. Note that the uncertainties are asymmetric and that the abundance scale is logarithmic. 	\label{fig:uncertall}}
\end{figure}

\begin{table}
        \centering
        \caption{Total production uncertainties for $p$ nuclides from the MC post-processing of the SN Ia model. The uncertainty factors shown for
variations up and down enclose a 90\% probability interval.}
        \label{tab:uncertall}
        \begin{tabular}{lrr}
                \hline
                Nuclide  &  Up &  Down \\
                \hline
${}^{74}\mbox{Se} $ & 1.020 & 0.988 \\
${}^{78}\mbox{Kr} $ & 1.018 & 0.974 \\
${}^{84}\mbox{Sr} $ & 1.022 & 0.986 \\
${}^{92}\mbox{Mo} $ & 1.034 & 0.978 \\
${}^{94}\mbox{Mo} $ & 1.041 & 0.967 \\
${}^{96}\mbox{Ru} $ & 1.034 & 0.976 \\
${}^{98}\mbox{Ru} $ & 1.030 & 0.978 \\
${}^{102}\mbox{Pd}$ & 1.044 & 0.970 \\
${}^{106}\mbox{Cd}$ & 1.034 & 0.975 \\
${}^{108}\mbox{Cd}$ & 1.035 & 0.974 \\
${}^{113}\mbox{In}$ & 1.170 & 0.896 \\
${}^{112}\mbox{Sn}$ & 1.030 & 0.963 \\
${}^{114}\mbox{Sn}$ & 1.040 & 0.969 \\
${}^{115}\mbox{Sn}$ & 1.346 & 0.845 \\
${}^{120}\mbox{Te}$ & 1.068 & 0.957 \\
${}^{124}\mbox{Xe}$ & 1.044 & 0.965 \\
${}^{126}\mbox{Xe}$ & 1.035 & 0.958 \\
${}^{130}\mbox{Ba}$ & 1.070 & 0.941 \\
${}^{132}\mbox{Ba}$ & 1.077 & 0.928 \\
${}^{136}\mbox{Ce}$ & 1.045 & 0.951 \\
${}^{138}\mbox{Ce}$ & 1.042 & 0.952 \\
${}^{138}\mbox{La}$ & 1.173 & 0.741 \\
${}^{144}\mbox{Sm}$ & 1.032 & 0.974 \\
${}^{152}\mbox{Gd}$ & 1.601 & 0.764 \\
${}^{156}\mbox{Dy}$ & 1.733 & 0.737 \\
${}^{158}\mbox{Dy}$ & 1.219 & 0.829 \\
${}^{162}\mbox{Er}$ & 1.988 & 0.848 \\
${}^{164}\mbox{Er}$ & 1.110 & 0.933 \\
${}^{168}\mbox{Yb}$ & 1.187 & 0.891 \\
${}^{174}\mbox{Hf}$ & 1.102 & 0.873 \\
${}^{180}\mbox{Ta}$ & 2.590 & 0.864 \\
${}^{180}\mbox{W} $ & 1.260 & 0.914 \\
${}^{184}\mbox{Os}$ & 1.170 & 0.880 \\
${}^{190}\mbox{Pt}$ & 1.195 & 0.858 \\
${}^{196}\mbox{Hg}$ & 1.060 & 0.936 \\
\hline
        \end{tabular}
\end{table}

\begin{table}
        \centering
        \caption{Same as Table \ref{tab:uncertall} but for radiogenic nuclides.}
        \label{tab:uncertradio}
        \begin{tabular}{lrr}
                \hline
                Nuclide  &  Up &  Down \\
                \hline
$\iso{Nb}{92}$  & 1.048 & 0.949 \\
$\iso{Tc}{97}$  & 1.040 & 0.968 \\
$\iso{Tc}{98}$  & 1.048 & 0.960 \\
$\iso{Sm}{146}$ & 1.067 & 0.943 \\

\hline
\end{tabular}
\end{table}

\begin{figure}
\includegraphics[width=\columnwidth]{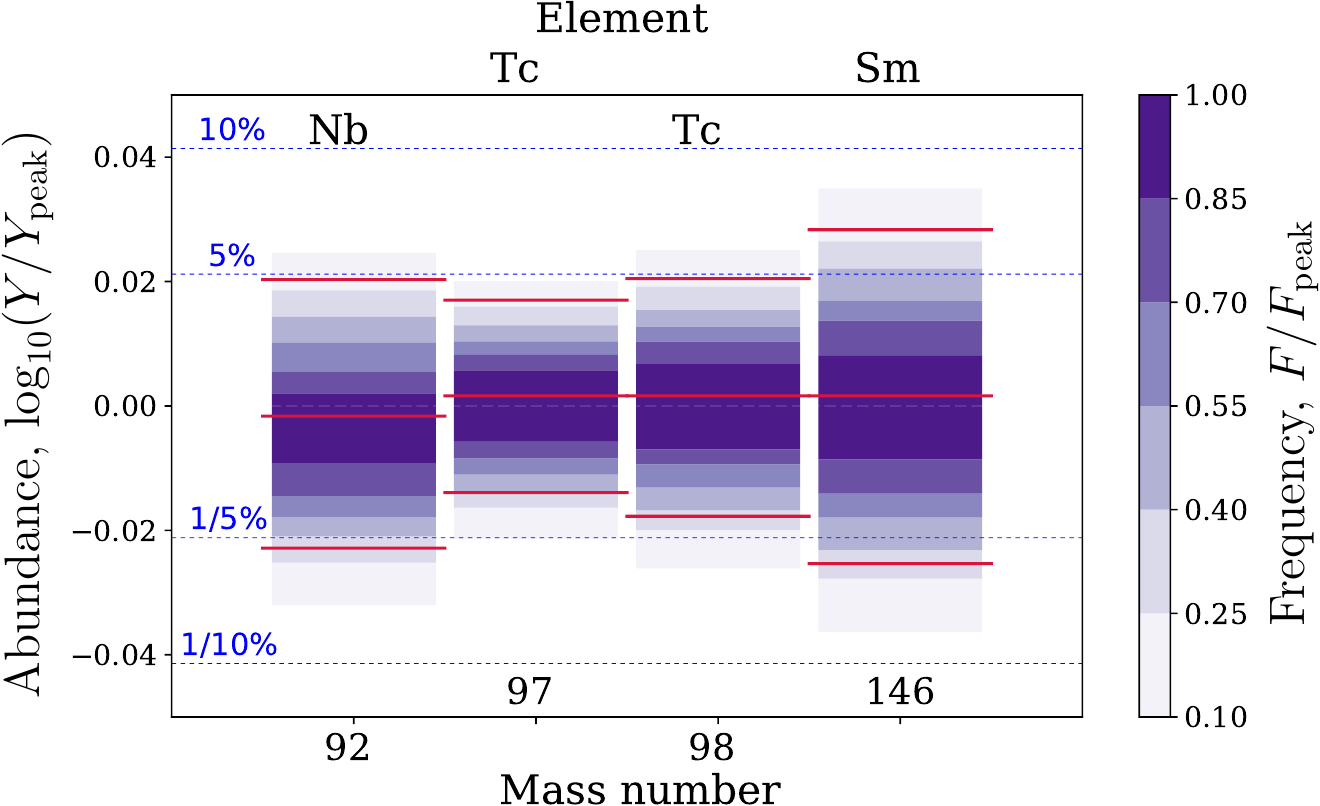}
\caption{Same as Fig.\ \ref{fig:uncertall} but for radiogenic nuclides. Note that the scale of the vertical axis is different from Fig.\ \ref{fig:uncertall}. \label{fig:uncertradio}}
\end{figure}

The combined production uncertainties of $p$ nuclei from all trajectories can be seen in Fig.\ \ref{fig:uncertall}. Shown are the abundance uncertainty distributions. The colour shade is the frequency $F$ of each abundance $Y$ normalized to the peak value of the distribution: $F_{\rm peak} \equiv F(Y_{\rm peak})$, where $Y_{\rm peak}$ is the peak position of $F(Y)$. Each distribution is asymmetric and not exactly a Gaussian or lognormal distribution, although the histogram has a continuous shape \citep[for further explanation, see Fig.~2 in][]{weakNobuya}. Horizontal red lines at each nuclide indicate $5\%$, $50\%$, and $95\%$ of the cumulative frequency and thus the interval between the $5\%$ and $95\%$ lines contains $90\%$ of the results. This interval is adopted as uncertainty in the final abundance. The numerical uncertainty value for each investigated nuclide is given in Table \ref{tab:uncertall}, in which the columns `Up' and `Down' correspond to the $Y(95\%)/Y_{\rm peak}$ and $Y(5\%)/Y_{\rm peak}$ values, respectively. Note that these comprise uncertainty factors which with an abundance has to be multiplied to obtain the uncertainty range or ``error bar''.

As in \cite{travradio} and \cite{pinccSn}, also the radioactive nuclides $^{92}$Nb, $^{97,98}$Tc, and $^{146}$Sm were included in the MC analysis. Their production uncertainties are given in Table \ref{tab:uncertradio} and shown in Fig.\ \ref{fig:uncertradio}.

The uncertainties are well below a factor of 2 (note the logarithmic scale in Fig.\ \ref{fig:uncertall}), with the exception of $^{180}$Ta. It should be noted that we did not explicitly follow $^{180g}$Ta and $^{180m}$Ta separately and thus the uncertainty in $^{180m}$Ta production may be even larger, depending on the unknown, actual equilibration between ground state and isomeric state \citep{rhhw02,mohr}. It can be noted that the uncertainties are generally larger from $^{152}$Gd on to higher mass numbers, with $^{162}$Er approaching an uncertainty of a factor of 2 at the upper limit. Incidentally, this is also the nuclear mass region where the calculations by \citet{travWD} exhibited problems to reproduce the solar abundance pattern for several nuclides. It has to be noted, however, that $^{113}$In, $^{115}$Sn, $^{138}$La, $^{164}$Er, $^{152}$Gd, and $^{180m}$Ta receive strong contributions from other processes than the $\gamma$ process \citep{neutrino,nemeth,arl99,travWD,p-review}. Therefore these nuclides should not be considered as (pure) $p$ nuclides although they were included in the classical list of $p$ nuclides by \cite{burb} and \cite{cam}.

The generally low uncertainty level can be understood by the fact that the SNe Ia trajectories cover a wider range of density--temperature combinations than, e.g., those of O/Ne layers in ccSNe (see Fig.\ \ref{fig:2groupsother}). This leads to a more complicated flow pattern without a well defined ``path'' when combining all trajectories. In such a flow, variations of a few rates do not greatly affect the general flow as they can easily be compensated by other reactions, and thus, this does not lead to an abundance variation.

Similar to the previous investigation of $p$ production in ccSNe by \citet{pinccSn}, key rates were identified by examining correlations between rate variations and abundance variations as described in Sec.\ \ref{sec:montecarlo}. Only one key rate was found\footnote{For the arrow notation see the remarks in \citet{pinccSn}.}, $\iso{Eu}{145} + \mbox{p} \leftrightarrow \gamma + {}^{146}$Gd (correlation factor $r=-0.72$), affecting the abundance of the radioactive nuclide $^{146}$Sm. This small number of key rates is consistent with the explanation for the low uncertainties as given above. Many different reactions in many different paths contribute to the abundance of a given nuclide and therefore also the total uncertainty is composed of the uncertainties of many rates and not just one or few key rates. This also explains the difference to \citet{pinccSn}, where many key rates were found. Those ccSN calculations yield a much narrower range of densities than found in WD explosions (see Fig.\ \ref{fig:2groupsother}) and thus also a better defined nucleosynthesis ``path''.

In our definition of key rates we adopted the same lower limit for the correlation factor of $|r|\geq 0.65$ as previously used in \citet{pinccSn}. Even with a lower limit of $|r|\geq 0.4$, we only found three more rates that are not key rates but are still of interest because they are contributing to the uncertainty: $\iso{Ge}{70} + \alpha \leftrightarrow \gamma + \iso{Se}{74}$ ($r=-0.46$) for $^{74}$Se, $\iso{Nd}{137} + \mbox{n} \leftrightarrow \gamma + \iso{Nd}{138}$ ($r=-0.51$) for $^{138}$Ce, and $\iso{Hf}{170} + \alpha \leftrightarrow \gamma + \iso{W}{174}$ ($r=-0.44$) for $^{174}$Hf. Additionally, the reaction $\iso{Eu}{145} + \mbox{p} \leftrightarrow \gamma + \iso{Gd}{146}$ appears again but also affecting $^{144}$Sm with a lower correlation ($r=0.50$) than for $^{146}$Sm. Except for this reaction, which appeared as a key reaction also in the 15 and 25 $M_\odot$ ccSN models, these reactions are different from the important rates identified in \citet{pinccSn}. Given the low uncertainties found for the nuclides affected by these reactions in SN Ia, however, even a highly accurate determination of these stellar rates will not alter the astrophysical conclusions, which rather are dominated by the uncertainties in the chosen astrophysical model.

\begin{figure}
\includegraphics[width=\columnwidth]{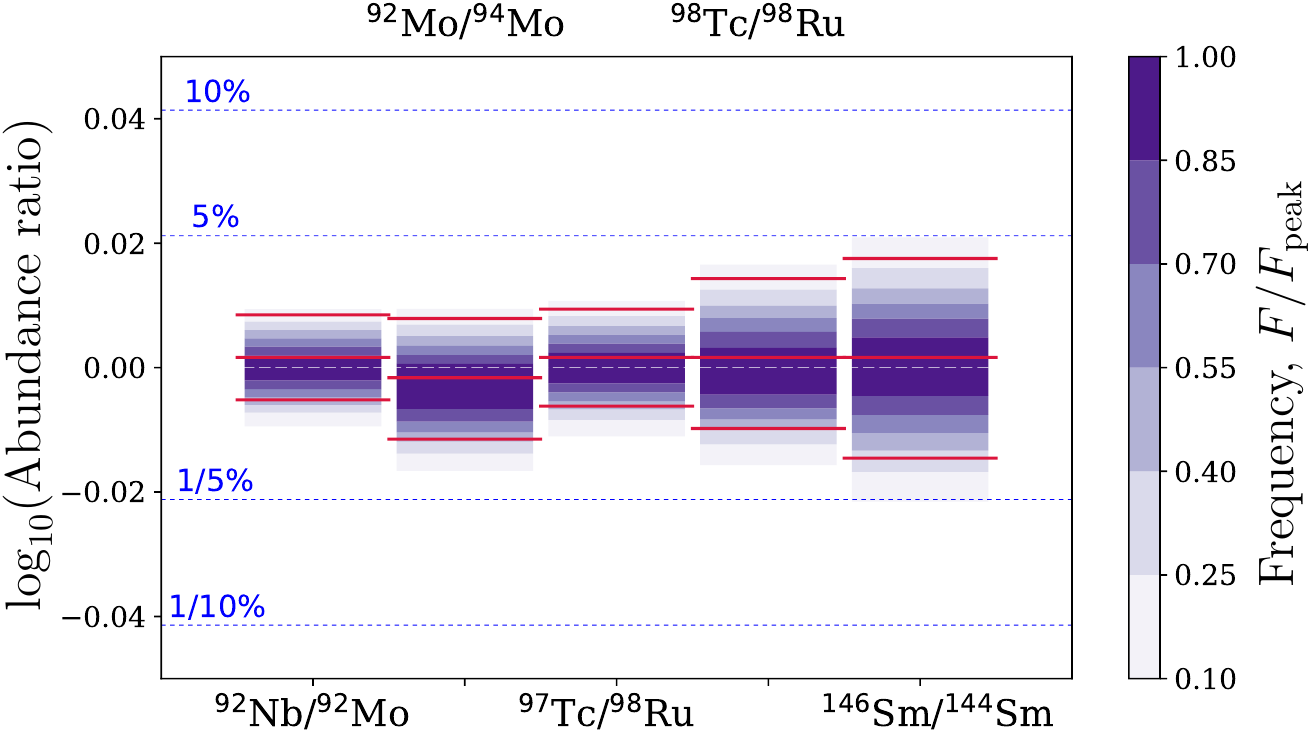}
\caption{Same as Fig.\ \ref{fig:uncertall} but for the logarithmic value ($\log_{10}$) of isotope ratios $Y(^{92}\mathrm{Nb})/Y(^{92}\mathrm{Mo})$, $Y(^{92}\mathrm{Mo})/Y(^{94}\mathrm{Mo})$, $Y(^{97}\mathrm{Tc})/Y(^{98}\mathrm{Ru})$, $Y(^{98}\mathrm{Tc})/Y(^{98}\mathrm{Ru})$, and $Y(^{146}\mathrm{Sm})/Y(^{144}\mathrm{Sm})$, normalized to the peak value of the distribution.  Note that the scale of the vertical axis is different from Fig.\ \ref{fig:uncertall}. \label{fig:isotoperatios}}
\end{figure}

For use in GCE calculations such as \citet{trav15}, we also determined uncertainty factors for the following abundance ratios: $Y(\iso{Nb}{92})/Y(\iso{Mo}{92})$, $Y(\iso{Mo}{92})/Y(\iso{Mo}{94})$, $Y(\iso{Tc}{97})/Y(\iso{Ru}{98})$, $Y(\iso{Tc}{98})/Y(\iso{Ru}{98})$, and $Y(\iso{Sm}{146})/Y(\iso{Sm}{144})$. These are shown in Fig.\ \ref{fig:isotoperatios}. As expected, the total uncertainties are tiny, at the same level as those for the light $p$ nuclides. Likewise, no key rate was found. The above mentioned reaction $\iso{Eu}{145} + \mbox{p} \leftrightarrow \gamma + \iso{Gd}{146}$ affects both $^{144}$Sm and $^{146}$Sm, although in opposite directions, and this is not sufficient to considerably change the $Y(\iso{Sm}{146})/Y(\iso{Sm}{144})$ ratio.

It should be noted that the above results concerning the production of individual nuclides as well as nuclide ratios are different from the conclusions drawn in \cite{travradio,trav15}. This is due to the fact that they followed a different approach and studied the impact of rates by individual variation of rates and identified important reactions by the sensitivity of abundances within a trajectory. As mentioned in Sec.\ \ref{sec:montecarlo}, an abundance sensitivity to a rate variation is not suited to quantify the importance of a rate \citep{omeg17} and such an approach also neglects the combined contributions of all trajectories to the final abundances and their uncertainties.

\subsection{Contributions from low- and high-density regions}
\label{sec:densityregions}

In order to better understand the origin of the uncertainties shown in Fig.~\ref{fig:uncertall} and to be able to draw conclusions also applicable beyond the specific SNe Ia model used here, it is interesting to separately consider tracers encountering different density regimes. As can be seen in Fig.~\ref{fig:2groupsother}, tracers fall into two groups, one at high densities of $4\times 10^8$--$10^9$~g\,cm$^{-3}$ (814 tracers) and one at lower densities of $10^6$--$10^8$~g\,cm$^{-3}$ (3810 tracers). Tracers with less than $5\times 10^6$~g\,cm$^{-3}$ were not found to contribute to $p$ abundances as their temperatures also remain too low. It can also be seen that the tracers in the high-density group do not cover the range of lower temperatures. They stem from the interior of the exploding WD, whereas the tracers in the low-density group are from layers close to the surface. Fig.~1 in \citet{travWD} shows the origin of the tracers in the simulation. In that figure, tracers in our high-density group are from the tracers in the WD interior and coloured mainly red (and some green).

\begin{figure}
\includegraphics[width=\columnwidth]{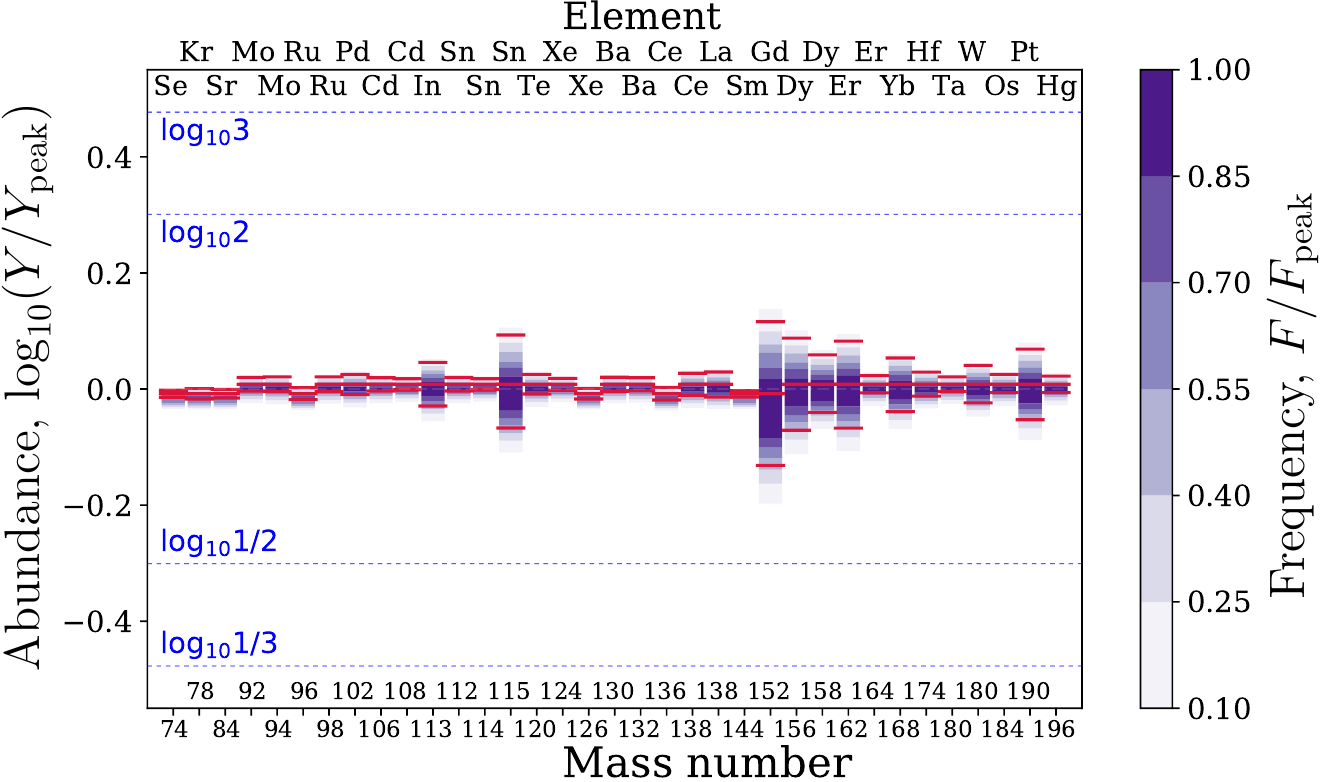}
\caption{Same as Fig.\ \ref{fig:uncertall} but considering only tracers in the low-density group. \label{fig:uncertlow}}
\end{figure}

\begin{figure}
\includegraphics[width=\columnwidth]{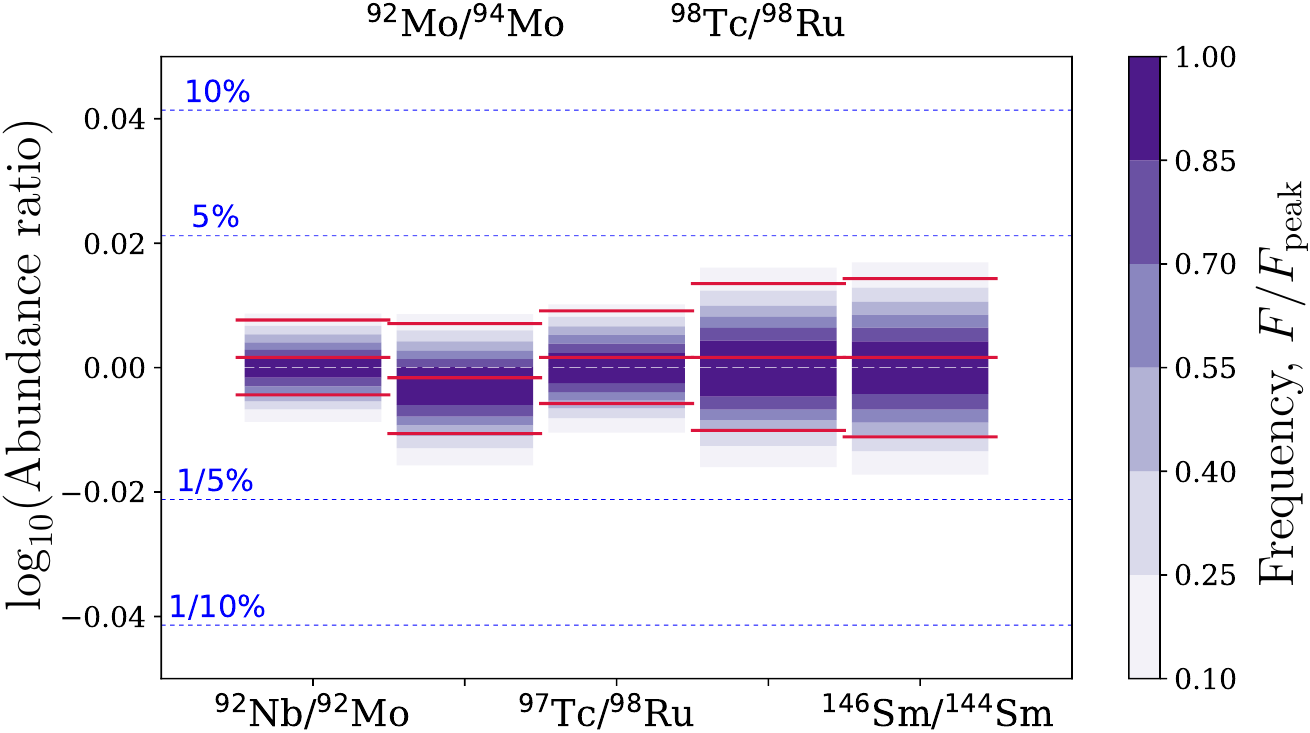}
\caption{Same as Fig.\ \ref{fig:isotoperatios} but considering only tracers in the low-density group. \label{fig:isotoperatios_lo}}
\end{figure}

\begin{figure}
\includegraphics[width=\columnwidth]{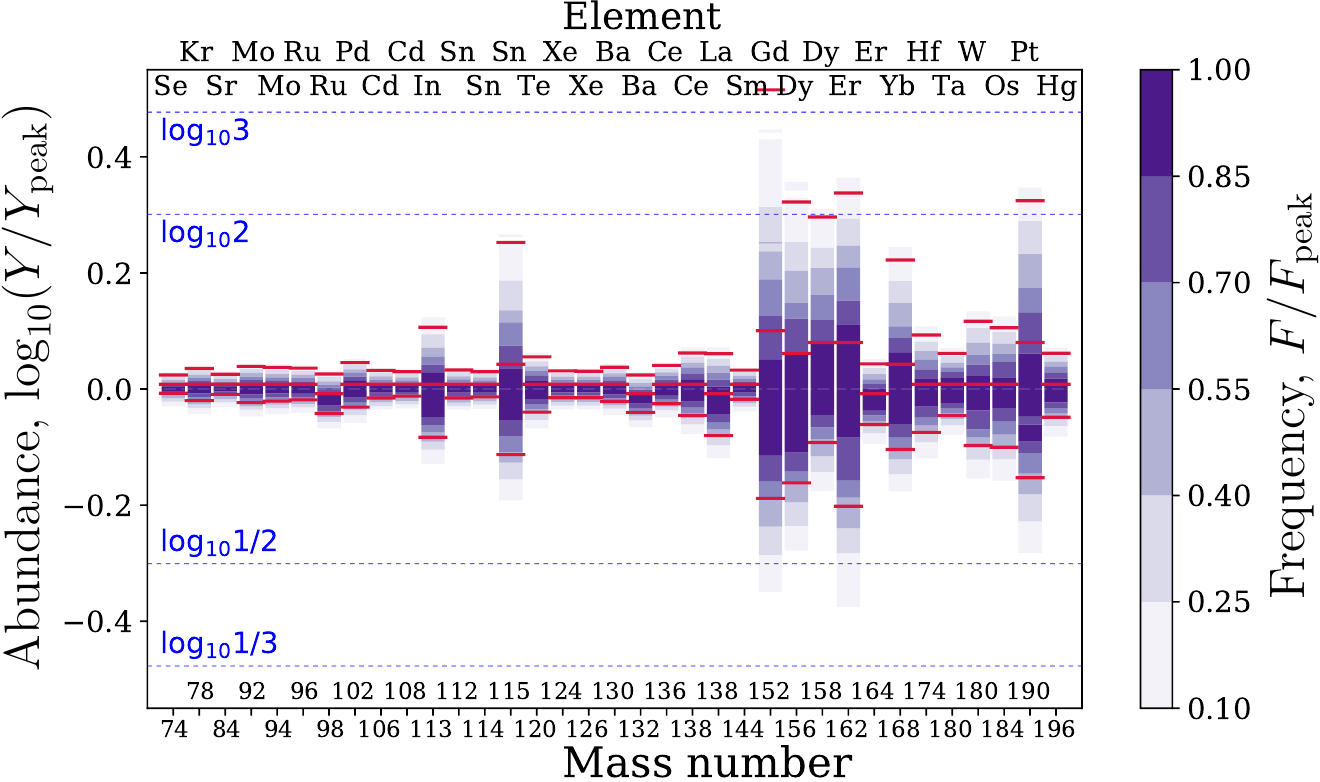}
\caption{Same as Fig.\ \ref{fig:uncertall} but considering only tracers in the high-density group. \label{fig:uncerthi}}
\end{figure}

\begin{figure}
\includegraphics[width=\columnwidth]{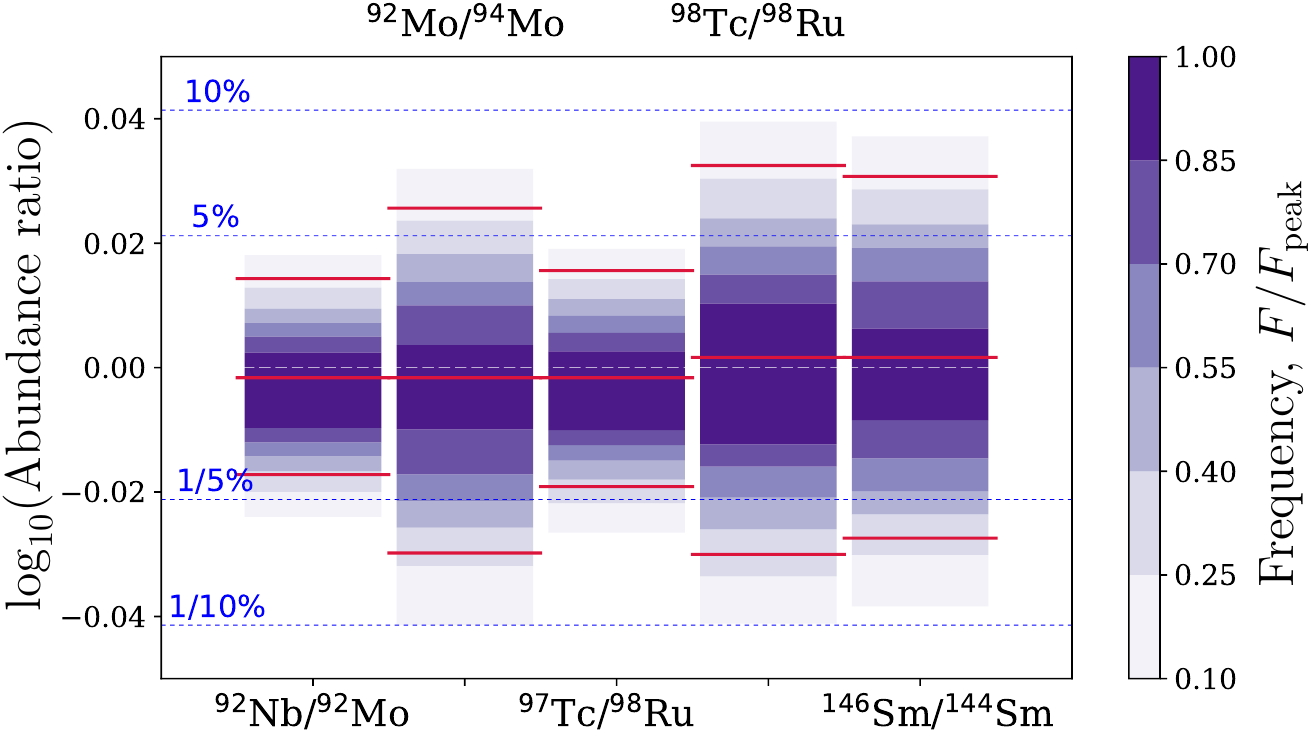}
\caption{Same as Fig.\ \ref{fig:isotoperatios} but considering only tracers in the high-density group. \label{fig:isotoperatios_hi}}
\end{figure}

In a further uncertainty analysis, the two groups were treated separately, as if belonging to two different models. The final abundance uncertainties of classical $p$ nuclei derived for the low- and high-density group are shown in Figs.\ \ref{fig:uncertlow} and \ref{fig:uncerthi}, respectively. Those of the production ratios
$Y(\iso{Nb}{92})/Y(\iso{Mo}{92})$, $Y(\iso{Mo}{92})/Y(\iso{Mo}{94})$, $Y(\iso{Tc}{97})/Y(\iso{Ru}{98})$, $Y(\iso{Tc}{98})/Y(\iso{Ru}{98})$, and $Y(\iso{Sm}{146})/Y(\iso{Sm}{144})$ are plotted in Figs.\ \ref{fig:isotoperatios_lo} and \ref{fig:isotoperatios_hi} for the low- and high-density group, respectively. It is apparent that the high-density group exhibits larger uncertainties whereas the uncertainties in the low-density group are smaller than the total uncertainties displayed in Fig.\ \ref{fig:uncertall}. Therefore the larger total uncertainties are mainly due to reactions in the high-density tracers.

Key rates were identified separately for the two groups with the same criteria as used for the total uncertainties and their key rates in Sec.\ \ref{sec:totalresults}. For the low-density group, five key rates were found: $\iso{Ba}{129} + \mbox{n} \leftrightarrow \gamma + \iso{Ba}{130}$ ($r=-0.68$) for $^{130}$Ba, $\iso{Ce}{137} + \mbox{n} \leftrightarrow \gamma + \iso{Ce}{138}$ ($r=-0.71$) for $^{138}$Ce, $\iso{Sm}{144} + \alpha \leftrightarrow \gamma + \iso{Gd}{148}$ ($r=-0.69$) for $^{146}$Sm, $\iso{Yb}{164} + \alpha \leftrightarrow \gamma + \iso{Hf}{168}$ ($r=-0.68$) for $^{168}$Yb, and $\iso{Pt}{186} + \alpha \leftrightarrow \gamma + \iso{Hg}{190}$ ($r=-0.70$) for $^{190}$Pt.

For the high-density group, seven key rates were identified: $\iso{Rb}{83} + \mbox{p} \leftrightarrow \gamma + \iso{Sr}{84}$ ($r=-0.65$) for $^{84}$Sr, $\iso{Cd}{105} + \mbox{n} \leftrightarrow \gamma + \iso{Cd}{106}$ ($r=-0.66$) for $^{106}$Cd, $\iso{Sn}{111} + \mbox{n} \leftrightarrow \gamma + \iso{Sn}{112}$ ($r=-0.77$) for $^{112}$Sn, $\iso{Ba}{129} + \mbox{n} \leftrightarrow \gamma + \iso{Ba}{130}$ ($r=-0.78$) for $^{130}$Ba, $\iso{Ce}{137} + \mbox{n} \leftrightarrow \gamma + \iso{Ce}{138}$ ($r=-0.72$) for $^{138}$Ce, $\iso{W}{176} + \alpha \leftrightarrow \gamma + \iso{Os}{180}$ ($r=-0.67$) for $^{180}$W, and $\iso{Pt}{186} + \alpha \leftrightarrow \gamma + \iso{Hg}{190}$ ($r=-0.67$) for $^{190}$Pt.

\begin{figure}
\includegraphics[width=\columnwidth]{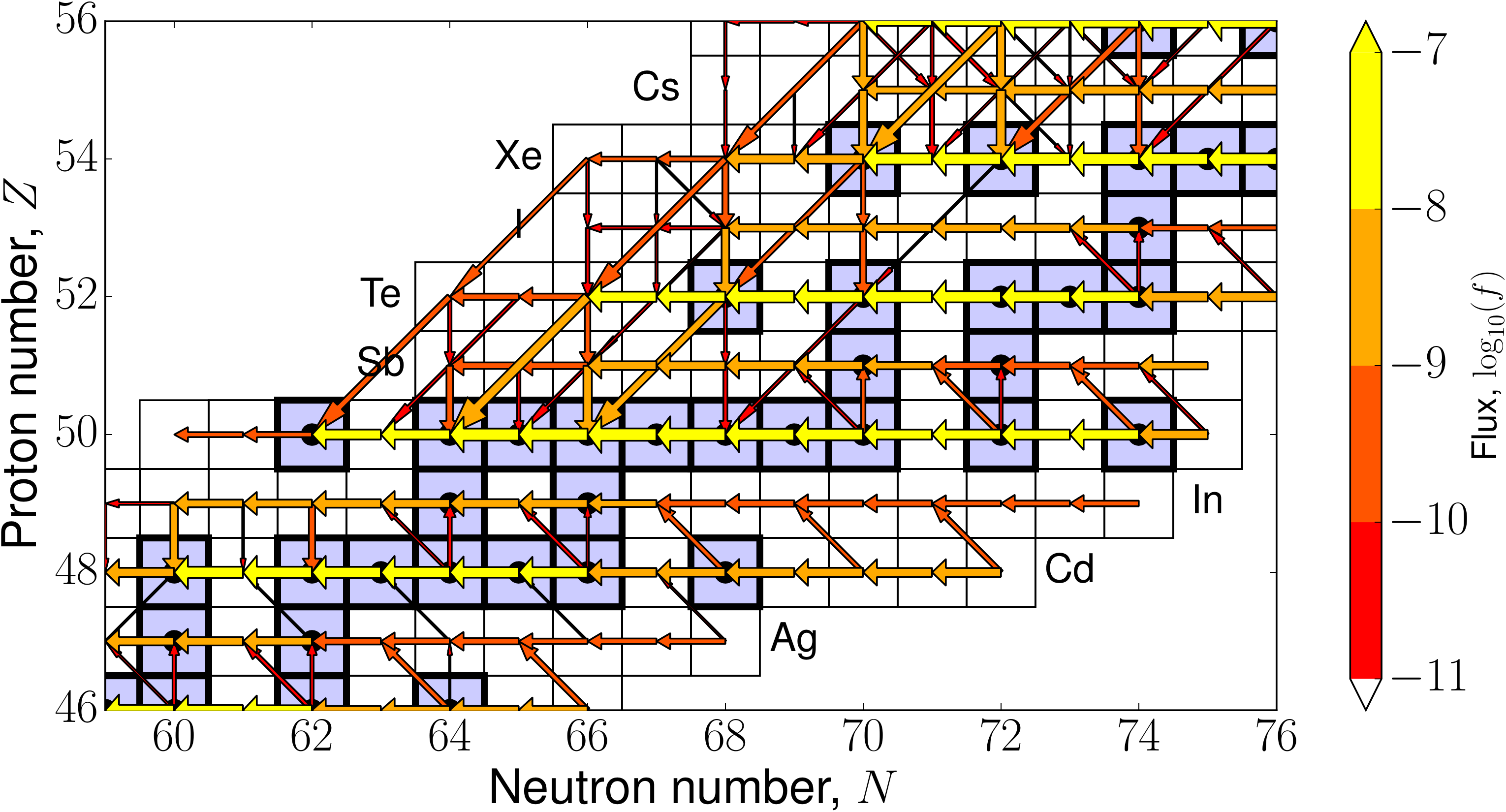}
\caption{Integrated reaction flows in the vicinity of Sn in a trajectory of the low-density group with peak temperature 2.8 GK. \label{fig:flowlowdensity}}
\end{figure}

\begin{figure}
\includegraphics[width=\columnwidth]{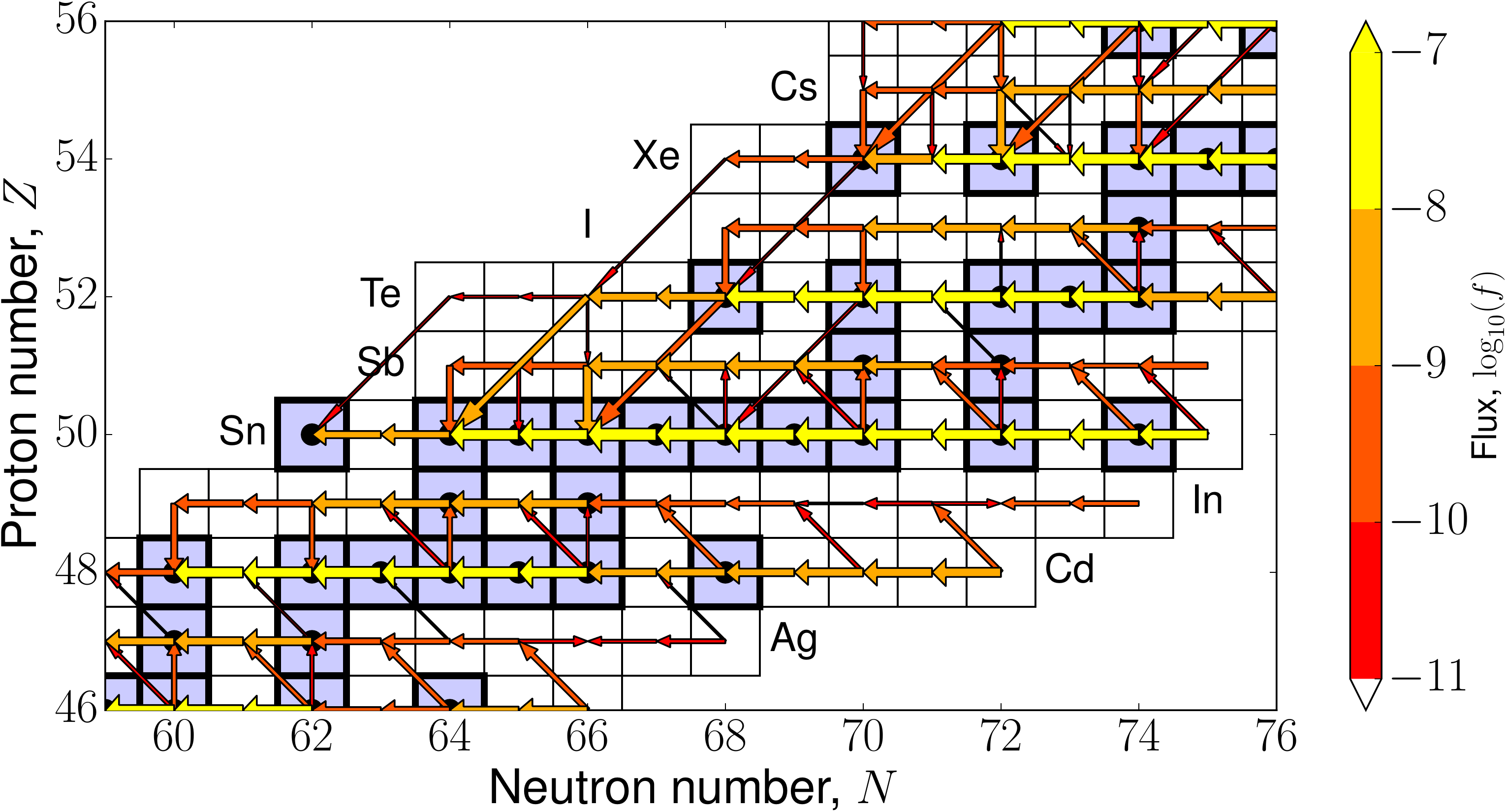}
\caption{Same as Fig.\ \ref{fig:flowlowdensity} but for a trajectory of the high-density group with the same peak temperature. \label{fig:flowhidensity}}
\end{figure}

Apart from the different number of tracers in each group, the physical differences between the two groups are the covered temperature and density ranges. The difference in density shifts the balance between captures and photodisintegrations, as the photodisintegration rate is not affected by the matter density. Higher density, on the other hand, accelerates capture rates relative to photodisintegrations which slows down the net flow (given by the difference between forward and reverse rates) towards the proton-rich side because of increased neutron captures. An example of this is seen when comparing Figs.~\ref{fig:flowlowdensity} and \ref{fig:flowhidensity}, showing flows in two trajectories from different density groups but with the same peak temperature. The flows in Fig.\ \ref{fig:flowlowdensity} extend further to the proton-rich side than the ones in Fig.~\ref{fig:flowhidensity}. Also the ($\gamma,\alpha$) downflows are increased, simply because these cannot be as strongly fed in the high-density regime. In the mass region of light $p$ nuclides, where Coulomb barriers are lower than at higher mass numbers. Furthermore, faster proton capture rates may become important in shifting the competition between ($\gamma$,p) and (p,$\gamma$) at higher density. The difference in covered temperature range is less important as there are a few low-temperature tracers also in the high-density group. Due to the small number of tracers experiencing low temperature at high density, there may be fewer flow alternatives and more emphasis on certain reactions affecting the heavy $p$ nuclides.

The combined effect of the above is well reflected in the difference in key rates identified in the two groups. Altough the overall number of key rates is small in both groups, there are more key rates on lighter and intermediate mass nuclides (involving reactions with neutrons and one with protons) in the high-density group.
The key rates affecting the abundance uncertainties of $^{130}$Ba, $^{138}$Ce, and $^{190}$Pt are the same in the two groups. None of the key rates in the two groups, nevertheless, appears as key rate or reaction of importance in the total uncertainties obtained from the combined analysis including both groups. This underlines that the importance of individual reactions strongly depends on available alternative flow paths that are more restricted when considering only a subset of trajectories.

The tracers used in this work are drawn from a specific calculation of a single model. Considering the two density groups, however, allow us to draw more general conclusions. For example, as can be seen in Fig.\ \ref{fig:2groupsother}, the W7 model of \citet{nom} coincides with the low-density group. Therefore it can be expected that similar uncertainties as shown in Fig.~\ref{fig:uncertlow} are obtained when using W7 trajectories. More generally, the obtained total uncertainties depend on the relative contributions of low- and high-density groups. Other 2D or 3D simulations of SNe Ia explosions may yield different tracer distributions between different density regimes. (As mentioned before, the required temperature range for $p$ nucleosynthesis is well defined by nuclear physics considerations and cannot vary between astrophysical models.) As long as they are close to the densities discussed here, the resulting total uncertainties can be estimated by an appropriately weighted combination of the uncertainties shown in Figs.~\ref{fig:flowlowdensity} and \ref{fig:flowhidensity}. Key rates, on the other hand, cannot be inferred from a simple combination as has become apparent in the above discussion.

\section{Summary and conclusions}
\label{sec:summary}

A comprehensive, large-scale MC study of the production of $p$ nuclei was performed using more than 4000 trajectories from a 2D model of a single-degenerate, thermonuclear supernova. Astrophysical reaction rates on several thousand target nuclides were simultaneously varied within individual temperature-dependent uncertainty ranges constructed from a combination of experimental and theoretical error bars. This allowed us to investigate the combined effect of rate uncertainties leading to total uncertainties in the final abundances. The large number of trajectories resulted in considerable computing requirements necessitating the use of HPC facilities.

In addition to the 35 classical $p$ nuclides, abundance uncertainties were also determined for the radioactive nuclides $^{92}$Nb, $^{97,98}$Tc, and $^{146}$Sm, important for GCE studies. Uncertainties of the abundance ratios $Y(\iso{Nb}{92})/Y(\iso{Mo}{92})$, $Y(\iso{Mo}{92})/Y(\iso{Mo}{94})$, $Y(\iso{Tc}{97})/Y(\iso{Ru}{98})$, $Y(\iso{Tc}{98})/Y(\iso{Ru}{98})$, and $Y(\iso{Sm}{146})/Y(\iso{Sm}{144})$ were also given. The total uncertainties found were well below a factor of two, despite of the contribution of a large number of unmeasured rates with large theory uncertainties, with the exception of $^{180m}$Ta and $^{162}$Er. These nuclides should not be expected to be produced purely by a $\gamma$ process, though. Key rates were identified using the correlation between rate variation and final abundance variation but only a single key rate was found, affecting the abundance of a nuclide with already very small uncertainty.

Although we used tracer particles from a specific 2D calculation as source for trajectories, more general conclusions can be drawn which are also applicable to other model calculations. Two groups of tracers coming from different locations in the WD, and thus experiencing different peak densities, were also analysed separately. Mass droplets in the right temperature range for $p$ nucleosynthesis may be distributed differently between different density groups in other 2D or 3D simulations of WD explosions. Nevertheless, the resulting total uncertainties in the production of $p$ nuclides can be assessed by appropriately combing uncertainties derived from the density groups studied here.

In conclusion, we found that the uncertainties stemming from uncertainties in the astrophysical reaction rates are small compared to the uncertainties arising from the choice of site, explosion model, and numerical treatment of the explosion hydrodynamics.

\section*{Acknowledgements}

We thank F. R\"opke for providing the DDT-a trajectories. We also thank M. Kusakabe and K. Nomoto for providing the trajectories of the W7 model included in Fig.\ \ref{fig:2groupsother}. This work has been partially supported by the European Research Council (EU-FP7-ERC-2012-St Grant 306901, EU-FP7 Adv Grant GA321263-FISH), the EU COST Action CA16117 (ChETEC), the UK STFC (ST/M000958/1), and MEXT Japan (Priority Issue on Post-K computer: Elucidation of the Fundamental Laws and Evolution of the Universe). GC acknowledges financial support from the EU Horizon2020 programme under the Marie Sk\l odowska-Curie grant 664931. CT acknowledges support from JINA-CEE. Parts of the computations were carried out on COSMOS (STFC DiRAC Facility) at DAMTP in University of Cambridge. This equipment was funded by BIS National E-infrastructure capital grant ST/J005673/1, STFC capital grant ST/H008586/1, and STFC DiRAC Operations grant ST/K00333X/1. DiRAC is part of the UK National E-Infrastructure. Further computations were carried out at CfCA, National Astronomical Observatory of Japan, and at YITP, Kyoto University. The University of Edinburgh is a charitable body, registered in Scotland, with Registration No.\ SC005336.








%
%


\bsp	
\label{lastpage}
\end{document}